\documentclass[final,3p,times]{elsarticle}
\usepackage{amssymb}
\usepackage{amsmath}
\usepackage{booktabs}
\usepackage[hidelinks]{hyperref}

\usepackage{xcolor}

\journal{Journal of Subatomic Particles and Cosmology}

\begin{document}
\begin{frontmatter}
\title{Hadron spectroscopy of QC$_2$D at finite density}

\author[a,b]{Kei Iida}
\author[c,d]{Etsuko Itou\fnref{speaker}}\ead{itou@yukawa.kyoto-u.ac.jp}
\author[d,e]{Kotaro Murakami}
\author[f]{Daiki Suenaga}

\fntext[speaker]{Speaker.}

\affiliation[a]{
  organization={Department of Liberal Arts,
                The Open University of Japan},
  city={Chiba},
  postcode={261-8586},
  country={Japan}
}

\affiliation[b]{
  organization={RIKEN Nishina Center, RIKEN},
  addressline={2-1 Hirosawa},
  city={Wako, Saitama},
  postcode={351-0198},
  country={Japan}
}

\affiliation[c]{
  organization={Yukawa Institute for Theoretical Physics,
                Kyoto University},
  addressline={Kitashirakawa Oiwakecho, Sakyo-ku},
  city={Kyoto},
  postcode={606-8502},
  country={Japan}
}

\affiliation[d]{
  organization={RIKEN Center for Interdisciplinary
                Theoretical and Mathematical Sciences
                (iTHEMS), RIKEN},
  addressline={2-1 Hirosawa},
  city={Wako, Saitama},
  postcode={351-0198},
  country={Japan}
}

\affiliation[e]{
  organization={Department of Physics,
                Institute of Science Tokyo},
  addressline={2-12-1 Ookayama},
  city={Megro, Tokyo},
  postcode={152-8551},
  country={Japan}
}

\affiliation[f]{
  organization={Kobayashi-Maskawa Institute for the Origin
                of Particles and the Universe,
                Nagoya University},
  city={Nagoya},
  postcode={464-8602},
  country={Japan}
}
\begin{abstract}
We investigate the chemical-potential dependence of hadron masses in two-color QCD (QC$_2$D) at low temperature using lattice Monte Carlo simulations. We calculate all possible meson and diquark correlation functions, including disconnected diagrams, on a $32^4$ lattice at $T\approx40\,\mathrm{MeV}$. In the hadronic phase, we obtain a meson mass hierarchy similar to that of three-color QCD. After the superfluid transition, the ordering changes: the $\sigma$ and $a_1$ channels become lighter, while the pion mass increases with density. 
We also examine the diquark source dependence of the Nambu--Goldstone mode of $U(1)_B$ symmetry breaking and find an almost vanishing mass in the superfluid phase after extrapolation to zero source. Finally, we compare correlation functions for chiral partners and discuss indications of partial chiral-symmetry restoration. This contribution is based on our recent study of hadron spectra in dense QC$_2$D~\cite{Iida:2026dmw}.
\end{abstract}
\begin{keyword}
Two-color QCD \sep Finite density \sep Hadron spectroscopy \sep Lattice QCD \sep Chiral symmetry
\end{keyword}
\end{frontmatter}

\section{Introduction}
How do hadron masses change in a dense medium? 
This question is closely related to the phase structure of QCD, since the condensates that characterize the vacuum also affect hadronic excitations.
An early proposal by Brown and Rho suggested that changes in the chiral condensate in a dense medium could be accompanied by changes in the masses of various hadrons~\cite{Brown:1991kk}.
Using QCD sum rules, Hatsuda and Lee predicted a decrease in the masses of light vector mesons in nuclear matter~\cite{Hatsuda:1991ez}.
However, subsequent QCD sum-rule analyses indicated that a reduction in the $\rho$ and $\omega$ masses is not inevitable in nuclear matter~\cite{Zschocke:2002mn,Steinmueller:2006id}.
At low temperature and high density, the formation of a diquark condensate may further modify the hadron spectrum.
Direct lattice Monte Carlo calculations in this region are, however, hindered by the sign problem in three-color QCD.

Two-color QCD (QC$_2$D) with two degenerate quark flavors has no sign problem even at a real quark chemical potential $\mu$ because of the pseudo-reality of the fundamental representation of $SU(2)$. 
This allows us to investigate the dense regime using the same Monte Carlo framework as at $\mu=0$. 
In our previous studies~\cite{Iida:2019rah,Iida:2024irv}, we determined the low-temperature phase structure mainly from the diquark condensate, the Polyakov loop and the quark number density. 
Near $\mu=m_\pi/2$, the hadronic phase changes into a superfluid phase with a nonzero diquark condensate. 
As $\mu$ increases further, the Bose--Einstein condensation (BEC) regime crosses over to a Bardeen--Cooper--Schrieffer (BCS) regime.

Having established this phase structure, we now turn to the hadron spectrum. 
Earlier lattice calculations found, for example, that the $\rho$ meson becomes lighter in the dense medium~\cite{Muroya:2002ry,Hands:2007uc}. 
To understand how the mass hierarchy changes more generally, we calculate the correlation functions for mesons and diquarks in all the allowed channels considered in our study, including disconnected contributions. 
In this article, we present our main results for the mass spectra and then discuss chiral symmetry through a comparison of chiral-partner correlators. 
The results are based on Ref.~\cite{Iida:2026dmw}.

\section{Hadron correlators}
Before presenting the numerical results, let us briefly describe the symmetries relevant to hadron spectroscopy in QC$_2$D. 
For $N_c=N_f=2$, the Pauli--G\"ursey symmetry, which allows transformations mixing quarks and antiquarks, enlarges the flavor symmetry to $SU(4)$ at $\mu=0$ in the massless limit.
A nonzero quark mass explicitly reduces it to $Sp(4)$, under which the three pions, the scalar diquark and the scalar antidiquark belong to the same multiplet~\cite{Kogut:2000ek}. 
At finite $\mu$, the hadronic phase preserves $SU(2)_V\times U(1)_B$. 
The diquark condensate spontaneously breaks $U(1)_B$ in the superfluid phase, while $SU(2)_V$ remains unbroken.

First, consider the hadronic phase, namely $\mu \lesssim m_\pi/2$ regimes, at zero temperature. 
For an operator carrying a quark number $n_O$, the temporal correlator can be expressed as
\begin{equation}
 C(\mu;\tau)=\sum_n |\langle0|\hat O|n\rangle|^2
                  e^{-(E_n-\mu n_O)\tau}.
 \label{eq:spectral}
\end{equation}
The decay rate at zero spatial momentum therefore gives
\begin{equation}
 m_{\mathrm{eff.}}(\mu)=m-\mu n_O,\qquad
 n_O=0,\,+2,\,-2
 \quad\text{for mesons, diquarks and antidiquarks, respectively}.
 \label{eq:shift}
\end{equation}
Here, $m$ is the mass at $\mu=0$. 
Thus, meson masses are independent of $\mu$, while diquark and antidiquark masses decrease and increase linearly, respectively.
This behavior is also predicted by the chiral perturbation theory (ChPT)~\cite{Kogut:2000ek}.
Here and below, we call the decay rate of the temporal correlator the effective mass, including the shift due to $\mu$.

In the superfluid phase, $m_\pi/2 \lesssim \mu$, on the other hand, mesons and diquarks are no longer distinguished by baryon number because $U(1)_B$ symmetry is spontaneously broken.
Meson and diquark operators with the same isospin and spin-parity $(I,J^P)$ can couple to the same energy eigenstates.
For the diquark operators used here, Fermi statistics allows the scalar, pseudoscalar and vector channels for $I=0$, and the axial-vector channel for $I=1$. 
Consequently, meson--baryon mixing is possible in the $\sigma$, $\eta$, $\omega$ and $a_1$ channels, but not in the $\pi$ and $\rho$ channels. Such mixing has also been investigated from the linear sigma model~\cite{Suenaga:2025sln}.

In this work, we calculate meson and diquark two-point functions separately. If both operators couple to the lightest state with the same $(I,J^P)$, their effective masses can become degenerate at sufficiently large temporal separation.

\section{Lattice setup and mass extraction}
As a lattice setup, we employ the Iwasaki gauge action and two degenerate Wilson fermions. 
The simulation parameters are
$(\beta,\kappa,N_s,N_\tau)=(0.800,0.159,32,32)$.
We use the gauge configurations generated in our previous study~\cite{Iida:2024irv} of the phase diagram and equation of state. 
Setting the pseudo-critical temperature at $\mu=0$ to $T_c=200\,\mathrm{MeV}$ gives $T\approx40\,\mathrm{MeV}$, or $0.19T_c$. At $\mu=0$, the pion mass and the pion-to-$\rho$ mass ratio are $am_\pi=0.620(1)$ and $m_\pi/m_\rho=0.813(1)$, respectively.
It gives the pion mass $m_\pi\approx738\,\mathrm{MeV}$ in physical unit. As for the chemical potential, we investigate $0\leq a\mu\leq0.75$. 
On these ensembles, the superfluid transition occurs at $a\mu_c=0.29$, and the BEC--BCS crossover is located around $a\mu=0.45$~\cite{Iida:2024irv}.

In the superfluid phase, we introduce a diquark source term,
\begin{equation}
 \mathcal L_j=-j\bigl(\bar\psi_1K\bar\psi_2^T-\psi_2^TK\psi_1\bigr),
 \qquad K=C\gamma_5T^2,
\end{equation}
into the Lagrangian, where $C$ is the charge-conjugation matrix and $T^2$ is the second Pauli matrix acting on color.
We use $aj=0$ for $a\mu\leq0.25$ and three finite sources, $aj=0.010,0.015,0.020$, for $a\mu\geq0.27$.
We first measure the correlators at each $j$ and then extrapolate the effective masses to $j\to0$ at fixed $\mu$.

To calculate the correlators, we consider both normal and anomalous quark propagators. 
The anomalous propagators connect two quark fields or two antiquark fields and contribute in the presence of the diquark source. 
We also evaluate disconnected diagrams, subtracting the vacuum contribution where necessary. 
For the connected parts, we use wall-type quark sources and local sinks. The disconnected traces are estimated with the noise method. 
We analyze 100--700 configurations for each parameter set and estimate statistical errors by the jackknife method.

We obtain local effective masses from neighboring temporal points using an exponential or a cosh form, and perform a constant fit in the plateau region. 
Note that at finite $\mu$, forward and backward propagations do not necessarily have the same decay rate, so we choose the fitting form according to the behavior of each correlator.
We then take the $j\to0$ limit, mainly using a linear function of $aj$. For the Nambu--Goldstone(NG) mode of $U(1)_B$ breaking, we additionally use
\begin{equation}
 am_{\mathrm{NG}}(aj)=c_0+c_1\sqrt{aj},
 \label{eq:ngextrap}
\end{equation}
which is motivated by the source dependence predicted by ChPT~\cite{Kogut:2000ek}.

\section{Results for the mass spectra}
\begin{figure}[h]
\centering
\includegraphics[width=.4\textwidth]{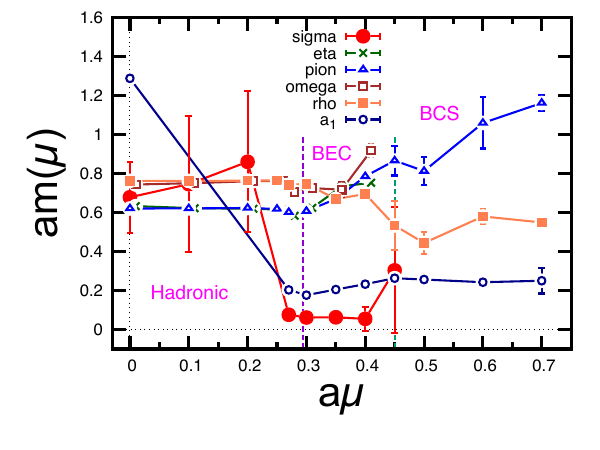}
\qquad
\includegraphics[width=.4\textwidth]{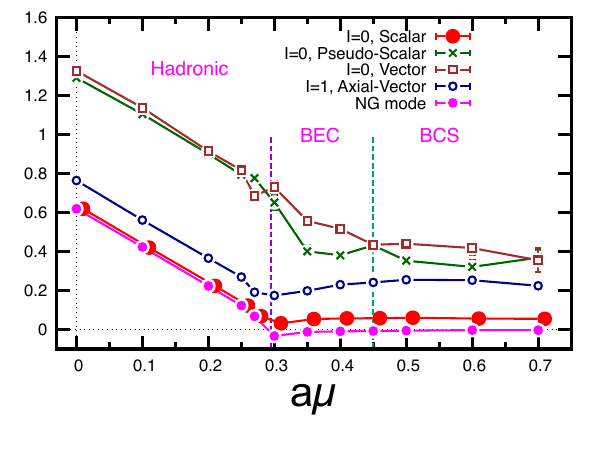}
 \caption{Masses extracted from mesonic operators (left) and from diquark and NG operators (right) as a function of $a\mu$. The labels indicate the hadronic, BEC and BCS regions of these ensembles. The results are from Fig.~26 of Ref.~\cite{Iida:2026dmw}.}
 \label{fig:spectra}
\end{figure}

Figure~\ref{fig:spectra} shows our results for the meson masses in the left panel and the diquark and NG masses in the right panel. 
Let us first focus on the mesons. 
In the hadronic phase, the masses are almost independent of $\mu$, as expected from Eq.~\eqref{eq:shift}. 
The pion is the lightest meson, and the $\rho$ and $\omega$ masses are nearly degenerate. 
After the superfluid transition, the ordering changes. 
We summarize the observed hierarchies as
\begin{align}
 \text{Hadronic phase:}\quad &
 m_\pi\lesssim m_\eta<m_\sigma\;\text{(noisy)}<m_\rho\sim m_\omega\ll m_{a_1},\notag\\
 \text{Superfluid phase:}\quad &
 m_\sigma\;\text{(noisy)}<m_{a_1}<m_\rho<m_\pi\sim m_\eta\;\text{(noisy)}
 \ll m_\omega\;\text{(noisy)}.
 \label{eq:hierarchy}
\end{align}
Here, ``noisy'' indicates channels for which the mass determination has large statistical uncertainties. 
The hierarchy in the hadronic phase resembles that in three-color QCD, whereas the $\sigma$ and $a_1$ channels lie below the pion in the superfluid phase.

The pion mass increases approximately linearly with $\mu$ after the transition, which qualitatively agrees with ChPT. 
In contrast, the $\rho$ mass decreases slightly. 
The latter behavior was also showed in earlier lattice results~\cite{Muroya:2002ry,Hands:2007uc, Murakami:2022lmq}. 
Since neither channel has a diquark partner with the same quantum numbers, these changes are not caused by the meson--baryon mixing discussed above.

On the other hand, why do the $\sigma$ and $a_1$ channels become so light? 
The $\sigma$ operator has the same quantum numbers as the scalar diquark sector, which contains the NG mode associated with spontaneous $U(1)_B$ breaking. 
Since we have not separated the mixed eigenstates, the light mass obtained from the $\sigma$ correlator can reflect this low-energy state. 
Similarly, the $a_1$ operator can mix with the relatively light $I=1$ axial-vector diquark. 
These observations suggest that meson--baryon mixing plays an important role in the changed hierarchy. 
In particular, including the disconnected contributions is essential for identifying the low-energy behavior of the $\sigma$ correlator.

The $\eta$ and $\omega$ results become noisier in the superfluid phase, but both suggest an increase in mass. 
For the $\eta$, one might expect a decrease at sufficiently high density if instanton effects are suppressed. 
In our previous calculation, however, the topological susceptibility remained nearly unchanged even in the BCS regime~\cite{Iida:2024irv}. Thus, the density range studied here does not show the suppression that would motivate such an expectation. 
Moreover, the $\eta$ and $\omega$ channels are both subject to meson--baryon mixing, so their individual mass eigenstates remain to be determined.

Next, we turn to the diquarks in the right panel of Fig.~\ref{fig:spectra}. 
Their masses decrease in the hadronic phase according to Eq.~\eqref{eq:shift}. 
Unlike the meson sector, the ordering is maintained across the transition:
\begin{equation}
 m_{NG}\lesssim m_{I=0,S}<m_{I=1,AV}<m_{I=0,PS}\lesssim m_{I=0,V},
 \label{eq:diquarks}
\end{equation}
At $\mu=0$, the scalar and axial-vector diquarks are related by the Pauli--G\"ursey symmetry to the pion and $\rho$ meson, respectively. 
This explains why these two diquark channels are relatively light. 
Antidiquark masses are not shown in Fig.~\ref{fig:spectra}, but they increase by $2\mu$ in the hadronic phase. 
In the superfluid phase, the diquark and antidiquark often have nearly degenerate masses, which come from  their coupling to the same energy eigenstate.

\begin{figure}[h]
 \centering
 \includegraphics[width=.4\textwidth]{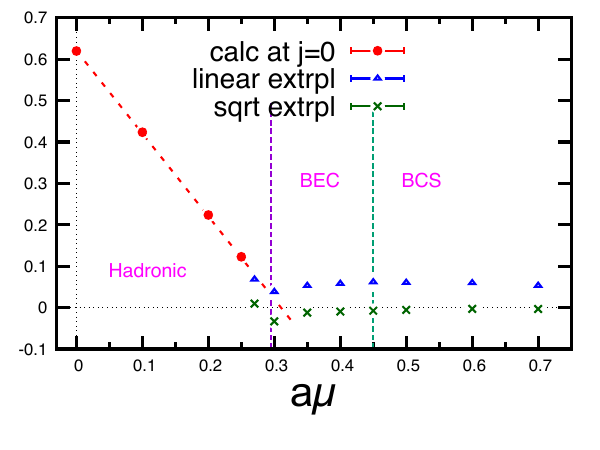}
\includegraphics[width=.4\textwidth]{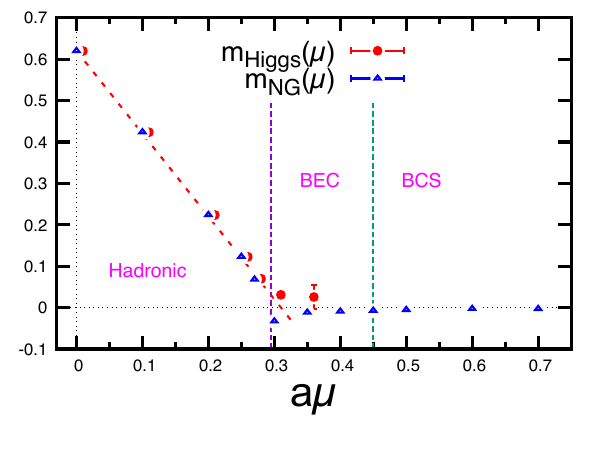}
 \caption{Results for the NG and Higgs masses in the $j \to 0$ limit, which is originally shown in Fig.~8 of Ref.~\cite{Iida:2026dmw}. Left: comparison of linear and square-root extrapolations of $j \to 0$ for the NG mass, together with direct $j=0$ results in the hadronic region. Right: NG and Higgs-channel results. At $a\mu\geq0.40$, the Higgs signal does not permit a reliable mass extraction.}
 \label{fig:ng}
\end{figure}
We further examine the NG and Higgs channels in Fig.~\ref{fig:ng}.
The left panel compares the NG masses obtained in the $j\to0$ limit using a linear extrapolation as well as the square-root form, Eq.~\eqref{eq:ngextrap}, motivated by the ChPT prediction.
The square-root extrapolation yields an almost vanishing mass in the superfluid phase, consistent with the NG mode associated with spontaneous breaking of $U(1)_B$.
Based on this comparison, we adopt the square-root extrapolation for our final NG masses, including those shown in Fig.~\ref{fig:spectra}.

The right panel compares these NG masses with the Higgs masses after taking the $j\to0$ limit.
In the hadronic phase, the two channels are degenerate, and their masses decrease toward zero as $\mu_c$ is approached.
In the superfluid phase, the NG mode remains nearly massless, whereas the Higgs correlator decays more rapidly and becomes increasingly noisy.
It suggests that the Higgs mode instead becomes heavier.
For $a\mu\geq0.40$, the Higgs signal is too noisy to extract a reliable mass.

\section{Chiral partners at high density}
Finally, let us examine chiral symmetry at high density.
In our earlier calculation, the chiral condensate was shown to decrease after the superfluid transition~\cite{Iida:2024irv}.
With Wilson fermions, however, the condensate requires an additive subtraction, and its nonzero value alone does not give a simple measure of chiral restoration.
We therefore compare the correlation functions of the chiral partners $\rho$ and $a_1$.

The vector and axial-vector operators,
$V_\mu^a=\bar\psi\gamma_\mu\tau^a\psi$ and
$A_\mu^a=\bar\psi\gamma_\mu\gamma_5\tau^a\psi$,
are related by axial transformations generated by $Q_5^a$:
\begin{equation}
 [Q_5^a,V_\mu^b]=i\epsilon^{abc}A_\mu^c,
 \qquad
 [Q_5^a,A_\mu^b]=i\epsilon^{abc}V_\mu^c.
\end{equation}
If chiral symmetry is restored, $Q_5^a|0\rangle=0$,
and invariance of the vacuum under these transformations,
together with the unbroken $SU(2)_V$ symmetry, implies
\begin{equation}
 \langle V_\mu^a(x)V_\nu^a(y)\rangle_0
 =
 \langle A_\mu^a(x)A_\nu^a(y)\rangle_0.
\end{equation}
This equality holds at each spacetime separation.

Here, the quark mass is finite, therefore, we examine whether these correlators
approach each other as an indication of a tendency toward chiral symmetry restoration.
We use the spatial components of the vector and axial-vector currents
to avoid the spin-zero contribution that can enter the temporal axial current.

\begin{figure}[h]
 \centering
 \includegraphics[width=.4\textwidth]{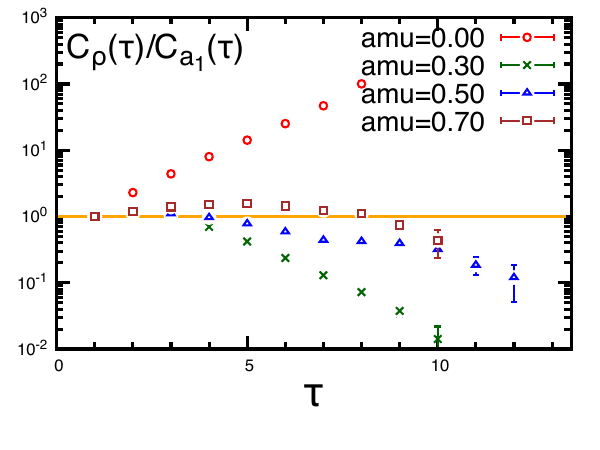} 
 \caption{Normalized $\rho$--$a_1$ correlator ratio at several chemical potentials, from Fig.~28 of Ref.~\cite{Iida:2026dmw}. Each curve is set to unity at $\tau=1$; the finite-density correlators are extrapolated to $j\to0$. At the largest $\mu$, the ratio stays near unity over a broader interval of temporal separation.}
 \label{fig:chiral}
\end{figure}
Figure~\ref{fig:chiral} shows the ratio of the $\rho$ and $a_1$ correlators at several chemical potentials.
Here, we extrapolate the finite-density correlators to $j\to0$ and then normalize the ratio at $\tau=1$.
At $\mu=0$, the ratio grows rapidly with $\tau$ because the $\rho$ is lighter than the $a_1$. 
In the superfluid phase, this ordering is reversed, and the ratio becomes smaller than unity for most temporal separations. 
As $\mu$ increases further, however, the ratio approaches unity over a wider range of $\tau$. At $a\mu=0.70$, we find approximate agreement up to $\tau\simeq8$.

This behavior suggests partial restoration of chiral symmetry at high density. 
The masses in Fig.~\ref{fig:spectra} are nevertheless still different: 
we extract them from a larger-$\tau$ region, where the slopes of the two correlators remain distinguishable. 
Thus, the tendency toward agreement at shorter separations does not imply degeneracy of the lowest masses. 
The normalization at $\tau=1$ also means that this plot compares the temporal dependence rather than the absolute amplitudes.

We have also compared the $\pi$ and $\sigma$ correlators in Ref.~\cite{Iida:2026dmw}. 
Their normalized ratio also moves closer to unity at higher density for $\tau\lesssim5$, while the errors at larger separations prevent a clear conclusion. 
Together with the $\rho$--$a_1$ comparison, these results indicate a tendency toward chiral restoration, although our quark masses are relatively heavy and chiral symmetry is explicitly broken by the Wilson action.

\section{Summary and outlook}
We investigated hadron spectra in dense two-color QCD at $T\approx40\,\mathrm{MeV}$. 
We calculated meson and diquark correlation functions, including disconnected diagrams. 
In the hadronic phase, the meson masses are almost independent of $\mu$.
The ordering of meson masses is similar to that in three-color QCD. 
On the other hand, the diquark masses decrease with the expected $-2\mu$ shift. 
After the superfluid transition, the meson hierarchy changes: the $\sigma$ and $a_1$ channels become light, the pion mass increases.
Furthermore, the $\rho$ mass decreases slightly.
The ordering of the diquark masses is almost maintained across the transition in the superfluid phase.
The NG mass becomes nearly zero correctly if we used the ChPT-motivated functional form in $j \to 0$ extrapolation. 
We also observed a tendency for chiral-partner correlators to approach one another at high density.
It suggests the restoration of chiral symmetry.

One remaining task is to disentangle meson--baryon mixing by calculating cross-correlators and solving a generalized eigenvalue problem. 
This will clarify which eigenstates dominate the light scalar and axial-vector channels. It would also be interesting to investigate hadron--hadron interactions in the superfluid phase. 
Our earlier study using the HAL QCD method addressed interactions in the hadronic phase~\cite{Murakami:2023ejc}; the changed mass hierarchy found here motivates extending that analysis to the dense medium.

\section*{Acknowledgments}
K.~I.\ is supported by JSPS KAKENHI with Grant No.\ 23K25864. 
K.~I.\ and E.~I.\ are supported by JSPS KAKENHI with Grant Number 25K01001. 
The work of E.~I.\ is supported by 
JST SQAI with Grant Number JPMJPF2221,  
JST CREST with Grant Number JPMJCR24I3.  
E.~I.\ and D.~S.\ are supported by JSPS KAKENHI with Grant Number 23H05439. 
D.~S.\ is also supported by JSPS KAKENHI No.~23K03377, No.~23H05439 and No.~25K17386.
K.~M.\ is supported in part by Grants-in-Aid for JSPS Fellows (Nos.\ JP22J14889, JP22KJ1870) and by JSPS KAKENHI with Grant No.\ 22H04917. 
This work is also supported by Program for Promoting Researches on the Supercomputer ``Fugaku'' (Simulation for basic science: from fundamental laws of particles to creation of nuclei) and (Simulation for basic science: approaching the new quantum era), and Joint Institute for Computational Fundamental Science (JICFuS), Grant Number JPMXP1020230411. 
This work is supported by Center for Gravitational Physics and Quantum Information (CGPQI) at Yukawa Institute for Theoretical Physics.

\bibliographystyle{elsarticle-num}
\bibliography{2color}
\end{document}